\documentclass[aps,prb,showpacs,twocolumn,floatfix]{revtex4}
\usepackage{graphicx}     
\begin{document}
\date{\today}
\title{ Electron-positron interaction in light elements\\
represented by atoms embedded in an electron gas}
\author{H. Stachowiak}
\author{E. Boro\'{n}ski}
\affiliation{Polish Academy of Sciences, W. Trzebiatowski Institute 
for Low Temperature and Structure Research\\
P.O.Box 1410, 50-950 Wroc{\l}aw 2, Poland}
\begin{abstract}
Mijnarends et al. 
[J. Phys. Condens. Matter {\bf 10}, 10383 (1998)]
contested the best existing calculations of positron
annihilation rates in jellium and crystal lattices, 
pointing in this way at deficiencies of existing theories of 
electron-positron interaction in 
these materials. In the present work the local enhancement factors due 
to $e^+-e^-$ interaction in Li, Be, B, C, N and O are computed in a consequent many-body
approach for core and as concerns lithium also for conduction electrons 
and compared to the results of existing approximations to this problem 
which avoid direct many-body calculations in metals, i.e. the local density, 
generalized gradient and weighted density 
approximations, as well as to experimental data. 
Conclusions about positron lifetime and
$e^+-e^-$ correlation energy are also presented. Suggestions concerning
annihilation rates in an electron gas agree with those of Mijnarends et al.
\end{abstract}
\maketitle
\section{Introduction}
The purpose of this work is a study of electron-positron interaction (EPI) 
in solids represented by an atom embedded in an electron gas. 
There are good reasons to believe that EPI in such a model gives valuable 
information on $e^+-e^-$ interaction. 
Indeed, in most calculations the local annihilation rates are assumed 
to depend only on the local electron density, in generalized gradient
approximation  also on its local gradient \cite{Barb}. 
In our work they are functionals of the whole
electron distribution in an atom while the neighboring atoms are
simulated by an averaged electron density and a constant positive background.

This model was used to compute the effect of EPI for light 
elements. However, the approach to electronic structure used in this work
(i.e. presenting the population of valence electrons by means of a single 
density amplitude in the form proposed by Gondzik and Stachowiak
\cite{Gondz}) seems to hold only for alkalis 
\cite{damp, Banach}.
So more or less 
reliable calculations of the enhancement of valence electrons in these
elements were performed only for lithium. As concerns other elements,
we had to limit ourselves to core electrons for which the density
amplitude is equivalent to the wave function.

The unknown effect of EPI on
annihilation characteristics has always been considered as an obstacle 
in interpretation of annihilation data \cite{Seeger}. Nevertheless 
positron annihilation found many applications in studies of the 
solid state and in other domains (including biology, medicine and
even marketing) and proved to give valuable
information on electronic and ionic structures 
\cite{Varenna, Manuel}.

EPI in many electron systems has also been
the subject of many investigations both theoretical and
experimental. A review of different approaches to this problem 
is given in Ref.~\onlinecite{Ishii}. 

In particular, the unknown effect of the 
interaction of the positron with core electrons (IPC) introduces
an uncertainty which is an obstacle in the interpretation 
of experimental data in terms of properties of valence electrons.

From the many attempts to estimate the effect of IPC (and EPI in general) 
on the annihilation 
data  let us mention the local density approximation (LDA) 
\cite{fin, Puska1},
the generalized gradient approximation (GGA) \cite{Barb}, the weighted
density approximation (WDA) \cite{Rub98} and others 
\cite{Carb, opt}.

In the works 
\cite{fin, Rub98} no attempt is performed to study
the behavior of electrons in solids in presence of the positron 
basing on the equations of quantum mechanics. They all benefit 
of the results of jellium calculations. GGA takes into account 
the inhomogeneity of the medium, but has to introduce 
a phenomenological constant in order to get agreement with 
experimental measurements of positron lifetimes in metals. This
agreement was contested as concerns other applications 
by Mijnarends et al. \cite{Mij} who 
performed two-detectors Doppler broadening measurements in Al.
WDA has the merit of enforcing a total charge of the screening 
cloud around the positron equal to one electronic charge. 

Among the many-body calculations the most advanced are those of Sormann
\cite{opt}. But even in them no attempt to reach self-consistency
is undertaken.

In the present work we will develop the approach proposed by Gondzik
and Stachowiak \cite{Gondz} in order to treat a positron in an 
electron gas and called mnemotechnically HNC (hypernetted-chain) 
following in that the work of Kallio et al. \cite{Kal}. The approach
of Ref.~\onlinecite{Gondz} is particularly simple, and leads to
reasonable results. So one can hope that generalizing it to real
solids will be easier than in the case of other approaches.
The basic equations used in the present work have been proposed in Ref.~\onlinecite{epling}
and the method of solving them for the anisotropic case (i.e. for the positron beyond
the center of symmetry) is presented in Ref.~\onlinecite{ncrev}.
This work constitutes an application of the methods elaborated in these last papers.

Indeed, some information about $e^+-e^-$ interaction in lithium
was already obtained along this line 
\cite{Kansas, epling}.
However, in these
works we were unable to solve the occuring integro-differential
equations in the two dimensions needed if the spherical symmetry is broken
(it will be broken for the positron outside the nucleus). In the 
meantime a method to solve such equations was elaborated and then
applied to lithium 
\cite{ncrev, Acta, munch}.

In this work we had to limit ourselves to light elements,
since at present we are able to deal only with core electrons in the 
{\it s} orbital state. So we will perform calculations only for Li,
Be, B, C, N and O.

The present state of the art can be characterized best basing
on the recent enlightening work of Mijnarends et al. \cite{Mij}. 
These authors interpret two-detectors Doppler broadening studies of Al
in the following way.

They assume three models of $e^+-e^-$ interaction which they label
LDA, LDA' and GGA. 
The first two models apply the local density approximation to
$e^+-e^-$ interaction using the results of electron gas theory,
in the first case those of Arponen and Pajanne (AP) \cite{Arp},
in the second case those of Lantto \cite{Lant}. The third case 
corresponds to correcting the results of the local density approximation
by adding a gradient correction to the enhancement
factors of AP according to the work of Barbiellini et al. \cite{Barb}.
Mijnarends et al. find that the LDA' model gives the best agreement 
with experiment, though a small correction by applying a lesser than
in Ref.~\onlinecite{Barb} gradient correction could maybe lead to 
some improvement.
These results illustrate the present state of knowledge of $e^+-e^-$
interaction both in an electron gas and in real metals. 
Comparison of GGA with
experiment performed in Ref.~\onlinecite{Mij} led to disagreement.

As concerns deviations of enhancement factors from the local density 
approximation in metal
lattices, this problem according to Ref.~\onlinecite{Mij} is of lesser
importance. But this conclusion follows from experimental considerations
and means that  a satisfactory theory of $e^+-e^-$ interaction in metal
lattices is not existing in spite of more than thirty years of research in
this direction. Moreover, the LDA' model is based on the calculations 
of Lantto which assume an oversimplified trial function of the Jastrow
type. This trial function neglects as well momentum dependence 
of $e^+-e^-$ scattering and dependence of $e^--e^-$ correlations 
on the distance from the positron - effects well established in physics
and included in other calculations 
\cite{Arp, Lowy, Rub, PHNC}.
\section{Computations}
{\bf The model}
\vskip2mm

Applying the approach of Gondzik and Stachowiak \cite{Gondz}
to the problem 
of positron screening in an electron gas, we describe the electronic 
structure of an atom with two electrons core embedded in jellium with 
two functions $\psi_1(r)$
and $\psi_2(r)$. $\psi_1(r)$ is the wave function of core electrons 
and $\psi_2(r)$ is the density amplitude of valence electrons.
$\psi_2^2(r)$ is equal to the density of conduction electrons in the model.
These two functions obey the appropriate Kohn-Shamlike equations 
(in atomic Hartree units which will be used throughout the paper):
\begin{eqnarray}
&[- \frac{1}{2} \nabla^2 + V({\bf r})] \psi_1({\bf r}) 
 = & E_1 \psi_1({\bf r}), \nonumber \\[3mm] 
&[- \frac{1}{2} \nabla^2 + V({\bf r})] \psi_2({\bf r})  = & 0 
\label{1}
\end{eqnarray}
where
\begin{widetext}
\begin{equation}
V({\bf r}) = - \frac{Z}{r} + 2 \int d{\bf r}' \frac{\psi_1^2({\bf r}')}
{\mid {\bf r} -{\bf r}'\mid} + \int d{\bf r}' 
\frac{\psi_2^2({\bf r}') - d({\bf r}')} {\mid {\bf r} -{\bf r}'\mid}
+ V_{HL} \{ 2 \psi_1^2 ({\bf r}) + \psi_2^2 ({\bf r})\}
-V_{HL}\{\rho_0\}.
\label{2}
\end{equation}
\end{widetext}
$Z$ is the charge of the nucleus, $d({\bf r})$ is
the distribution of the positive charge in the electron gas.
It is equal
\begin{equation}
d(r)=
\cases
{\rho_0 & {\rm for \hspace{3mm}}  $r>R_{WS}$, \vadjust{\kern2mm}\cr
 0 &  {\rm for \hspace{3mm}} $r<R_{WS}$ 
}
\label{15}
\end{equation}
where
\begin{equation}
\rho_0 = D \frac{3 (Z-2)}{4 \pi (R_{WS})^3}.
\label{16}
\end{equation}
$R_{WS}$ is the radius of the Wigner-Seitz sphere. Eqs. (\ref{1}) -- (\ref{2})
limit us to elements having a two-electrons core. 
$V_{HL}\{\rho\}$ is the 
Hedin-Lundqvist exchange-correlation correction for an electron gas 
of density $\rho$ \cite{HL}. 
The Lagrange multiplier $E_2$ which should occur on the 
right-hand side of the second equation (\ref{1}) is normalized to zero 
by the last term in the formula (\ref{2}), while $E_1$ 
is the energy eigenstate of core electrons.
\begin{figure}
\includegraphics[width=80mm]{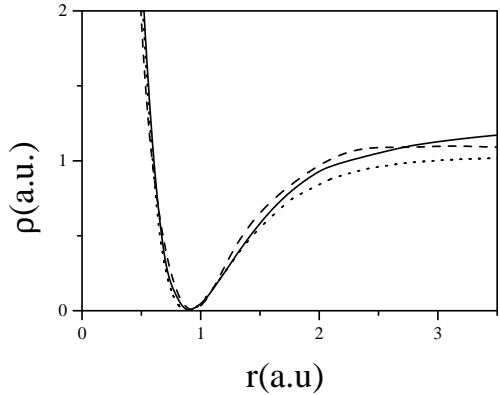}
\caption{ \label{Fig1c} Comparison of the density of valence electrons for a lithium
atom in an electron gas - full curve for 
$D=1.24$ and dotted curve for $D=1$ 
and the analogous density (averaged over direction) for 
metallic lithium in HNC approximation \protect\cite{damp}  dashed curve.}
\end{figure}  

We admit, of course, that from the elements treated in this work
only lithium can be described as above 
\cite{damp, Banach}.
But since
core electrons are only weakly affected by valence electrons, the 
way of presenting the electronic structure of these lasts is of
little importance for our purpose. In fact in this work we used 
different ways of approximating the electronic structure of valence 
electrons, but this had a negligible influence on IPC.

$D$ was chosen in order to satisfy the obvious normalization condition 
\begin{equation}
4 \pi \int_0^{R_{WS}} r^2 dr \psi_2^2(r) =Z-2.
\label{D}
\end{equation}
This requirement could be satisfied for lithium where the value 
$D=1.24$ leads to satisfying Eq. (\ref{D}) and in beryllium where
the value $D=1.1$ was obtained.
In Fig.~\ref{Fig1c} the distribution of
valence electrons $\psi_2^2(r)$ is shown for $D=1.24$ and compared 
to the distribution of valence electrons in metallic lithium obtained
in HNC formalism \cite{damp} after averaging over directions.
It is visible that the value $D=1.24$ reproduces better the electron 
distribution in the immediate neighborhood of the atom. 
\vskip2mm
{\bf Introduction of the positron}
\vskip2mm
We benefit of the result obtained according to the theory of liquids
by Kallio et al. \cite{Kal} 
and concerning the role of the light mass of the positron in EPI.
This suggests to describe 
the electronic structure of the model in presence of a positron 
at ${\bf r}_p$ by means of the equations
\begin{displaymath}
\left[ - \frac{1}{2} \nabla^2 + V(r) 
+ \frac{1}{2} W({\bf r}_p,{\bf r}) \right] \chi_1({\bf r}_p,{\bf r})
= E_1(r_p) \chi_1({\bf r}_p,{\bf r}),
\end{displaymath}
\begin{equation}
\left[ - \frac{1}{2} \nabla^2 + V(r) 
+ \frac{1}{2} W({\bf r}_p,{\bf r}) \right] \chi_2({\bf r}_p,{\bf r})=0
\label{p1}
\end{equation}
where the screened electron-positron potential $W({\bf r}_p,{\bf r})$
is defined as
\begin{equation}
W({\bf r}_p,{\bf r})= - \frac{1}{\mid {\bf r} -{\bf r}_p \mid}
+W_p({\bf r}_p,{\bf r}) + W_{xc}({\bf r}_p,{\bf r}) .
\label{Bp7}
\end{equation}
Here
\begin{widetext}
\begin{equation}
W_p({\bf r}_p,{\bf r})= 2 \int d{\bf r}' \: \frac{\chi_1^2({\bf r}_p,
{\bf r}') - \psi_1^2({\bf r}')}{\mid {\bf r} -{\bf r}' \mid}
+ \int d{\bf r}' \: \frac{\chi_2^2({\bf r}_p,
{\bf r}') - \psi_2^2({\bf r}')}{\mid {\bf r} -{\bf r}' \mid} ,
\label{Bp8p}
\end{equation}
\begin{equation}
W_{xc}({\bf r}_p,{\bf r})= 
V_{HL} \{ 2 \chi_1^2 ({\bf r}_p,{\bf r}) + \chi_2^2 ({\bf r}_p,{\bf r})\}
- V_{HL} \{ 2 \psi_1^2 ({\bf r}) + \psi_2^2 ({\bf r})\} .
\label{Bp9p}
\end{equation}
\end{widetext}
$\chi_i$ indicate the form of the functions $\psi_i$ in presence 
of the positron.
The positron distribution is described, of course, by the positron 
wave function.

Since the dependence of IPC on the perturbation of valence electrons 
by the positron is negligible, the equation for $\chi_1$ can be solved 
separately. However, we consider as a better approximation to solve
exactly the equations (\ref{p1}) for the easy problem of the positron
on the nucleus and to compute $\chi_1$ from the equation 
\begin{eqnarray}
\left[- \frac{1}{2} \nabla^2 +V^1({\bf r}) +
\frac{1}{2} W^1({\bf r}_p,{\bf r})\right] 
\chi_1({\bf r}_p,{\bf r})  & \nonumber \\
= E_1(r_p) \chi_1({\bf r}_p,{\bf r}). &
\label{BC6}
\end{eqnarray}
Since the positron in metals is always screened by valence electrons, 
we found that when computing $\chi_1$
it would be appropriate to freeze the valence electrons in the state they 
acquire for the positron on the nucleus. This will allow to determine the potentials 
$V^1$ and $W^1$  in Eq. (\ref{BC6}). This equation will be solved afterwards
for all values of ${\bf r}_p$.

Let us compute the solution of Eq. (\ref{p1}) for the positron 
on the nucleus, assuming that the core is frozen. We get in this way 
the potential
\begin{eqnarray}
Q({\bf r}) = &\int d{\bf r}' \frac{\chi_2^2(0,r')- \psi_2^2(r')}
{\mid {\bf r} - {\bf r}' \mid} + V_{HL} \{2 \psi_1^2(r) + \chi_2^2(0,r)\} \nonumber
\samepage \\ 
&- V_{HL} \{2 \psi_1^2(r) +\psi_2^2(r) \} .
\label{cor1}
\end{eqnarray}
The potentials in Eq. (\ref{BC6}) can now be written in the form
\begin{equation}
V^1(r) = V(r) + \frac{1}{2} Q(r),
\label{cor2}
\end{equation}
\begin{displaymath}
W^1({\bf r}_p,{\bf r}) = -\frac{1}{\mid {\bf r} - {\bf r}_p \mid }
+2 \int d{\bf r}' \frac{\chi_1^2({\bf r}_p,{\bf r}')- \psi_1^2(r')}
{\mid {\bf r} - {\bf r}' \mid} 
\end{displaymath}
\begin{equation}
+ V_{HL} \{2 \chi_1^2({\bf r}_p,{\bf r}) + \chi_2^2(0,r)\}
- V_{HL} \{2 \psi_1^2(r) +\chi_2^2(0,r) \}.
\label{cor3}
\end{equation}

\vskip2mm
{\bf Solution of the equations}
\vskip2mm

From test calculations it follows that the effect of positron interaction with
conduction electrons depends only slightly also on the polarization of
the core by the positron. So the equation for $\chi_2$ can be solved 
separately. 

The functions $\chi_i$ are presented in the form
\begin{equation}
\chi_i({\bf r}_p,{\bf r}) = A_i(r_p) e^{-\alpha s}+ \tau_i ({\bf r}_p,{\bf r})
\label{w13}
\end{equation}
where the function
\begin{equation}
\tau_i ({\bf r}_p,{\bf r})
=\sum_{n=0}^{\infty} \varphi_n^i(r_p,r) P_n(\cos \vartheta)
\label{tau}
\end{equation}
is devoid of the cusp 
occuring in $\chi_i$ at the positron. $P_n(\cos \vartheta)$ are
Legendre polynomials where $\vartheta$ is the angle between ${\bf r}$
and ${\bf r}_p$, $s=\mid {\bf r} - {\bf r}_p \mid$. 
From the assumption 
\begin{equation}
\tau_i({\bf r}_p,{\bf r}_p) = 0
\label{tau0}
\end{equation}
(the simplest one but not necessarily the only possible) 
it follows that $\alpha = 1/2$.

We have to limit ourselves for technical reasons 
to two terms in the expansion (\ref{tau}).
This leads to some problems especially for higher values of $r_p$.
They are discussed in more detail in Ref.~\onlinecite{ncrev}. 

The contact densities of electrons on the positron are provided 
by the $A_i(r_p)$ coefficients. An important information is also
contained in the energy eigenvalue $E_1(r_p)$.
Since $2E_1(r_p)$ will enter the positron Hamiltonian
as a contribution to the positron potential, this quantity can be
interpreted as effective attraction (of chemical character) between the 
positron and the nucleus due to collectivization of core electrons.
\section{Results}
We were able to obtain numerical values for local annihilation
rates of positrons in lithium. 
The results are shown on the figures. 
For other light elements we were able to study the effect of the interaction
of the positron with atomic cores. 

We call LDA the local density approximation,
mentioning each time what results are used for describing the properties of
a homogeneous electron gas: the ones of Gondzik and Stachowiak \cite{Gondz}
or PHNC \cite{PHNC} (eventually the formula of Boro\'nski and Nieminen (BN)
\cite{BN}). Since the interactions of the positron
with conduction and with core electrons can be to a large degree 
considered as independent, we introduce the term {\it local partial
density approximation} (LPDA) which means that only core electrons 
or only conduction electrons (depending which electronic state is
studied) contribute to the local density.

Note that the enhancement factors following from Eqs. (\ref{p1})
are approximate as concerns conduction electrons since they neglect
the difference of scattering on the positron (and also on the nucleus)
by different electronic states.
This is why it is appropriate to compare them while using LPDA with
the results of the Gondzik-Stachowiak approach. 
This problem does not occur in the case of core electrons.

The enhancement factors for core ($i=1$) and for valence electrons ($i=2$)
are given by the square of the enhancement amplitude $w_i$ 
which is defined as 
\begin{equation}
w_i({\bf r}_p) = \frac{A_i({\bf r}_p)}{ \psi_i({\bf r}_p)}
\label{enha}
\end{equation}

\begin{figure}
\includegraphics[width=53mm]{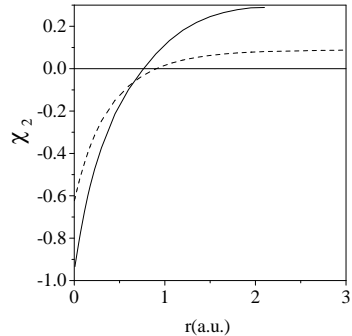}
\caption{\label{Fig1} The density amplitude of valence electrons in lithium 
as seen
by positrons $\chi_2(r_p,r_p)$ (full curve) compared to the true 
density amplitude $\psi_2(r_p)$ (dashed curve).}
\end{figure}

On Fig.~\ref{Fig1} the density amplitude as seen by positrons 
$\chi_2(r_p,r_p)$ is compared to the real density amplitude $\psi_2(r_p)$.

The enhancement amplitude of conduction electrons calculated according to 
(\ref{enha})
agrees quite well with LPDA predictions at $r_p=0$. Also for $r_p$
much bigger than the position of the node in $\psi_2$ it approaches 
the LPDA value. Note, however, that for big $r$'s its value is higher
than expected on ground of the local density approximation. 
We can interpret this effect as a result of 
increasing electron density between (approximately) the position of the node
in $\psi_2$ and $r < 2.5$. This result is confirmed by the recent
calculations of Boro\'nski and Stachowiak \cite{BS2001} who study 
the possibility of applying a grid method for problems of that kind.
A similar effect has been observed for vacancies \cite{ncrev}
where a maximum of the enhancement factor occurs in the region of increasing
electron density. In the present case we rather attribute this effect 
to the lack of
periodicity of the model. In the intermediate region $A_2(r_p)$ is
greatly affected by the displacement of the node  
(from $0.9$ to $0.76$ a.u.) due to interaction with the positron.We consider
this effect as very important, impossible to obtain using other approaches to
EPI. Such an effect has been observed experimentally by Chiba in MgO 
\cite{Chiba}.

\begin{figure*}
\includegraphics[width=145mm]
{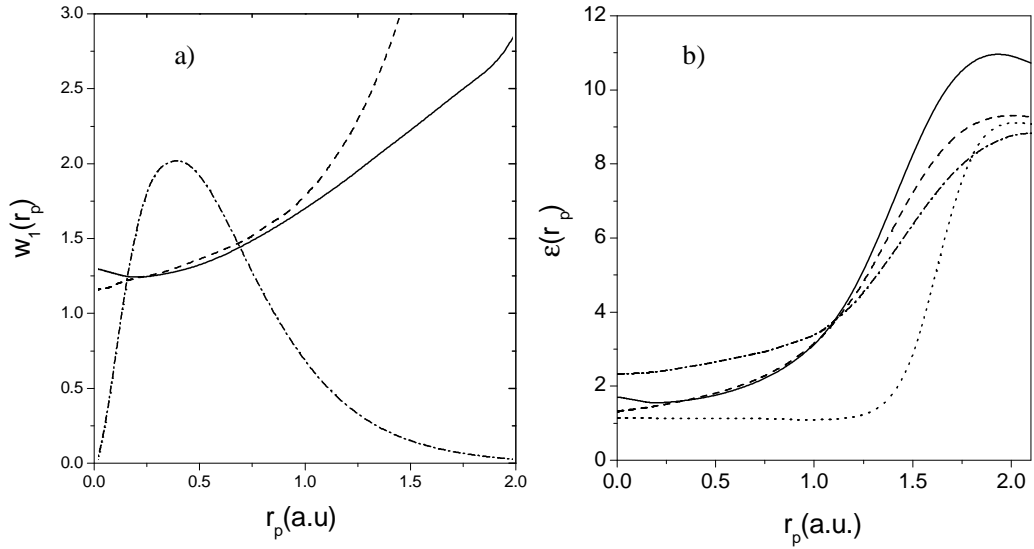}
\caption{a) The enhancement amplitude of core electrons in Li
according to this work (full curve). The dashed curve shows the LPDA 
prediction. Additionally, the core electron distribution 
$4\pi r^2 \psi_1^2(r)$ is plotted in arbitrary units (dashed-dotted curve).\\
b) Effective enhancement $\varepsilon (r_p)$ inside an atom of lithium 
embedded in an electron gas calculated 
in this work (solid curve)
and according to LDA (dashed curve), GGA (dotted curve) and WDA 
(dashed-dotted curve).}
\label{Fig2}
\end{figure*}

The enhancement amplitude for core electrons is presented in 
Fig.~\ref{Fig2}a for Li. 
Remark that in the immediate vicinity of the nucleus the enhancement 
amplitude is a little bigger than LPDA predictions, while for
higher values of $r$ it falls below the LPDA curves. 

Fig.~\ref{Fig2}b shows that the effective enhancement for lithium
defined as
\begin{equation}
\varepsilon(r_p) =  w_{eff}^2(r_p) = \frac{2 A_1^2(r_p)
+{\tilde A}_2^2(r_p)}{2 \psi_1^2(r_p) + \psi_2^2(r_p)} 
\label{92}
\end{equation}
reproduces quite well LDA predictions. 
${\tilde A}_2(r_p)$ has been
obtained from $A_2(r_p)$ by renormalizing it, taking account of the 
increasing error while $r_p$ increases when solving Eqs. (\ref{p1})
and the deviation of the approximation of Ref.~\onlinecite{Gondz} from 
the exact value assumed to correspond to PHNC \cite{PHNC}.

Of course, the local enhancement factor presented in Fig.~\ref{Fig2}a as 
well as the 
local enhancement of conduction electrons  
influence not only the positron lifetime but also angular correlation 
results.

As concerns the energy of $e^+-e^-$ correlation $E_{+-}$, 
\begin{figure}
\includegraphics[width=80mm]
{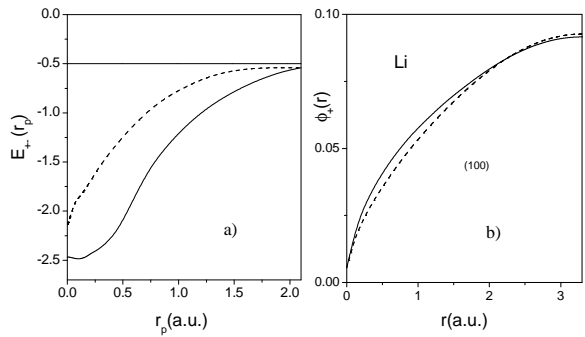}
\caption{\label{Fig3} a) Comparison of $E_{+-}(r_p)=2E_1(r_p)+E_v$ 
in Li (full curve) 
and the energy of electron-positron correlation  according to LDA 
(dashed curve). 
The energies have been normalized in such a way as to coincide at 
$r_p = 2.1$ a.u. \\
b) The positron density in Li along the (100) direction. 
The dashed
curve was obtained while using the $e^+-e^-$ correlation energy in LDA
approximation. The full curve follows from including the effect 
of nucleus-positron attraction due to collectivization of core electrons.}
\end{figure}
one can assume that
conduction electrons give to it a constant (independent of $r$)
contribution $E_v$, while the contribution of core electrons is provided 
by $2E_1(r_p)$. 
In Fig.~\ref{Fig3}a we compare $E_{+-}=2E_1(r_p)+E_v$
to the appropriate predictions of LDA for the energy of $e^+-e^-$
correlation. The jellium values were chosen according to PHNC
\cite{Bor98}. 

In Fig.~\ref{Fig3}b we compare the positron wave function in lithium 
calculated using 
the LDA approximation for the $e^+-e^-$ correlation potential to the wave 
function
obtained when the effect of correlation was estimated 
according to our results.

 We found also total annihilation rates for these two cases.
\begin{figure}
\includegraphics[width=55mm]
{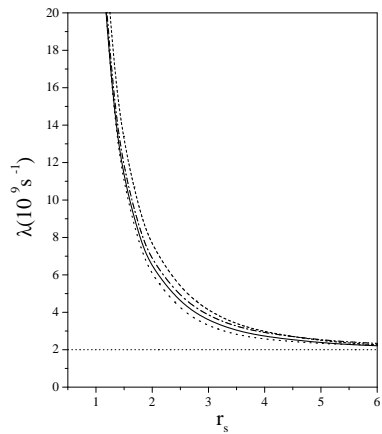}
\caption{Comparison of different formulas for the positron annihilation
rate $\lambda (r_s)$ in an electron gas (dashed curve -- Gondzik-Stachowiak 
\protect\cite{Gondz}, dashed-dotted curve -- 
Barbiellini {\it et al} \protect\cite{Barb}, 
solid curve -- PHNC \protect\cite{PHNC}, 
dotted curve -- Boro\'nski-Nieminen \protect\cite{BN}).}
\label{Fig4}
\end{figure}
We used the LDA approximation for the enhancement factor estimating
that in the light of Fig.~\ref{Fig2}b it is quite justified. 
In the first case, when we applied the LDA approximation to the 
PHNC enhancement factor (using the PHNC formula \cite{PHNC}) and to 
the $e^+-e^-$ correlation potential (with the formula from 
Ref.~\onlinecite{Bor98})
we obtained a total annihilation rate of $3.63$ $10^9/s$.
The electron densities were 
 calculated on base of the FLAPW (full potential linearized plane wave) 
method 
(the corresponding numerical code was WIEN95 \cite{Wien}).   

The bigger penetration
into the core region by the positron as shown in Fig.~\ref{Fig3}b  results in
slightly higher annihilation rates than in LDA because of high 
electron densities inside the core.
Thus, for the potential calculated according to this work,
the corresponding rate was $3.68$ $10^9/s$. 

The calculations performed with the BN enhancement
\cite{BN}  and the LDA correlation potential
 give a rate of $3.36$ $10^9/s$.
On the other hand, the calculations performed with the BN enhancement 
and our correlation potential give a rate of $3.42$ $10^9/s$, what is
accidentally in perfect agreement with experiment.
The experimental annihilation rate for lithium is equal $3.436$ $10^9/s$
\cite{gent}.

When calculating the positron wavefunction we used the numerical 
program by M. Puska \cite{priv} based, in principle, 
on superposition of atomic densities. 
 We changed it slightly, however, in order to have the possibility
to use electron densities and potentials calculated within the FLAPW method.
Moreover, we generalized it on an arbitrary
 non-LDA $e^+-e^-$ correlation potential. Note that Puska himself obtained an 
annihilation rate of $3.28$ $10^9/s$ basing on electron densities 
calculated according to the LMTO-ASA   (linearized muffin-tin orbital - atomic
sphere approximation) approach. Sormann and \v{S}ob \cite{ssmunch} remarked
already that different ways of computing band structure lead to predicting
different positron annihilation characteristics.
\section{Conclusions} 
In this work we tried on the simple example of lithium to find in a
consequent many-body way (by solving in a non spherically symmetrical 
surrounding the appropriate nonlinear integro-differential equation
derived on ground of the theory of liquids) deviations of local positron 
annihilation rates in a metal lattice from the local density approximation.
We came to the conclusion that in lithium no drastic deviation from LDA 
occurs as concerns the annihilation rate.
However, as concerns valence electrons, we observe a displacement 
of the node from 0.9 a.u. to 0.76 a.u., similar to that detected 
by Chiba in MgO \cite{Chiba}. We found also a deviation of the energy 
of electron-positron correlation from the local density approximation 
usually assumed in calculations. We found instead a term in the energy 
which we interpret as attraction of chemical character between the
nucleus and the positron due to collectivization of core electrons. 

So we computed the total positron annihilation rate in lithium using the
local density approximation  for the local annihilation rates. We used
the numerical ATSUP code of Puska \cite{priv} for calculating positron
wave functions, however, unlike Puska, we performed band structure
calculations using the FLAPW code labelled WIEN95 \cite{Wien}. Moreover
we added to the positron potential the term describing the positron-nucleus
attraction as obtained in our calculations. As concerns jellium 
annihilation rates, we used the formula of Boro\'nski and Nieminen
\cite{BN}. In this way we obtained perfect agreement with
experimental values and with the conclusions of Mijnarends {\it et al.} \cite{Mij} 
concerning the LDA' model. 

We confess that we are not very happy about this result. The formula of
Boro\'nski and Nieminen is the most frequently used in calculations 
trying to reproduce experimental data. Obviously it was found to give
the best agreement. However, it is based on the calculations of Lantto
which assume a Jastrow type trial function. This function neglects 
the momentum dependence of electron-positron scattering and the influence 
of the positron on electron-electron correlations. So we would have 
more confidence in calculations which take these effects, so well 
established in physics, into account. However, the annihilation rates
obtained from them are obviously too high (Fig.~\ref{Fig4}).
So we have to accept the Boro\'nski-Nieminen formula as an expression
poorly explained theoretically, but describing pretty well 
experimental data.

Among other results contained in the paper let us mention calculations 
of positron-nucleus attraction for Be, B, C, O and N due to collectivization
of core electrons, of core enhancement for these elements 
(Appendix \ref{appA})
and description
of the distribution of the electronic cloud screening the positron 
in different positions (Appendix \ref{appB}).
\acknowledgments
We are greatly indebted  to Dr. M.J. Puska for making accessible to us 
his numerical code ATSUP for calculating positron wave functions.
\appendix
\section{Positron-core interaction in light elements}
\label{appA}
For light elements heavier than beryllium the approach described 
by Eqs. (\ref{16}) and (\ref{D}) failed. Indeed, 
in boron it yielded $D=0.59$ and collapsed in carbon where
the condition (\ref{D}) needed a value of $D$ equal $2.5\cdot 10^{-3}$.
This, of course, is connected with the properties of valence 
electrons which are no longer itinerant. For this reason the condition 
(\ref{D})   was replaced for boron, carbon, nitrogen and oxygen 
by the condition
\begin{equation}
4 \pi \int_0^{cR_{WS}} r^2 dr \psi_2^2(r) 
+ \frac{4}{3} \pi \rho_0 (1-c^3) (R_{WS})^3 = Z-2
\label{Dc}
\end{equation}
while in Eq. (\ref{Dc}) the role of $R_{WS}$ is played now by $c R_{WS}$.
The normalization condition for $\psi_1$ is the usual one.
Note that the dependence of IPC characteristics on the values 
of $D$ and $c$ in Eqs.~\ref{16} and (\ref{Dc}) is negligible. 
Our calculations show that the enhancement amplitude for core
electrons in Be, B, C, O and N is even lower from the LDA curve
than the corresponding figure for Li. In general, however, its
behavior is quite similar to that shown in Fig.~\ref{Fig2}a.
Unfortunately, the effective enhancement for those elements
could not be computed for reasons explained in Section II.

As concerns the energies of positron-electron correlation for
Be, B, C, O and N the effect of core electron collectivization
leads to an attraction between the positron and the nucleus.
The effect is bigger for increasing atomic mass of the element. 
\section{The density amplitude $ \Large \chi_2({\bf r}_p,{\bf r})$}
\label{appB}
\begin{figure}
\includegraphics[width=80mm
]
{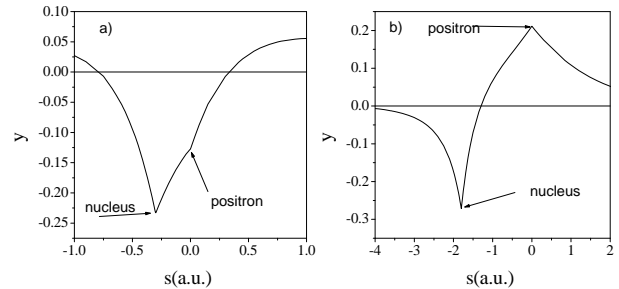}
\caption{\label{Fig6c}
$\chi_2({\bf r}_p,{\bf s}+{\bf r}_p) - \psi_2({\bf s}+{\bf r}_p)$
in Li for $r_p$ equal a) 0.3 b) 1.8~. ${\bf s}$ is taken along the line
connecting the nucleus and the positron.}
\end{figure}

\begin{figure}
\includegraphics
[width=80mm
]
{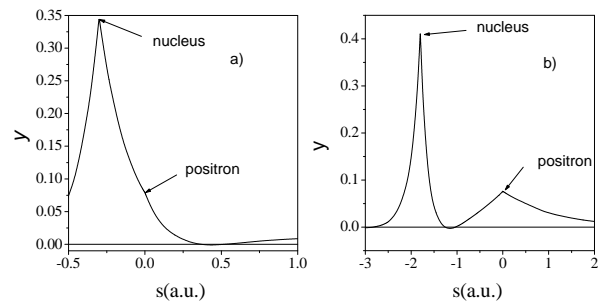}
\caption{\label{Fig7c}
The screening cloud distribution in Li around a positron at $r_p$ 
equal a) 0.3, b) 1.8 . ${\bf s}={\bf r}-{\bf r}_p$ is taken along the line 
connecting the nucleus and the positron.}
\end{figure}

Fig.~\ref{Fig6c} shows the change of the density amplitude 
$\chi_2({\bf r}_p,{\bf r}) - \psi_2({\bf r})$ for ${\bf r}$ along
the line connecting the nucleus and the positron and $r_p$ equal
a) 0.3 and b) 1.8~. Fig.~\ref{Fig7c} shows the behavior of the screening cloud
$\chi_2^2({\bf r}_p , {\bf r}) - \psi_2^2({\bf r})$ for the same values
of ${\bf r}$ and ${\bf r}_p$. One should remember that the corresponding density 
amplitude changes sign when crossing the node, and this feature persists
in the presence of the positron. Moreover, the node is shifted by the
positron. 
It is striking that the highest density
of the screening cloud occurs close to the nucleus even for the positron 
well beyond the node region of the density amplitude. This suggests that our
figures are less reliable for the positron in the interstitial region,
 since the screening cloud extends over several atomic cores and these facts
 are not included~in~our~model. \\[1mm]
%

%
\end{document}